\documentclass[prd,twocolumn,floatfix,reprint,aps,superscriptaddress]{revtex4}

\usepackage{amsfonts}
\usepackage{mathrsfs}
\usepackage{amsmath}% needed for subequations
\usepackage{color}
\usepackage{graphicx}
\usepackage{bm}% bold maths
\usepackage{amssymb}
\usepackage{xspace}
\usepackage{epstopdf}
\usepackage{dcolumn}% Align table columns on decimal point
\usepackage{longtable}
\usepackage{multirow}
\usepackage[colorlinks=true, letterpaper=true, pdfstartview=FitV, linkcolor=blue, citecolor=blue, urlcolor=blue]{hyperref}

\begin{document}

\title{Two-dimensional nodal-loop half metal in monolayer MnN}

\author{Shan-Shan Wang}
\affiliation{Research Laboratory for Quantum Materials, Singapore University of Technology and Design, Singapore 487372, Singapore}

\author{Zhi-Ming Yu}
\affiliation{Research Laboratory for Quantum Materials, Singapore University of Technology and Design, Singapore 487372, Singapore}

\author{Ying Liu}
\affiliation{Research Laboratory for Quantum Materials, Singapore University of Technology and Design, Singapore 487372, Singapore}

\author{Yalong Jiao}
\affiliation{Research Laboratory for Quantum Materials, Singapore University of Technology and Design, Singapore 487372, Singapore}

\author{Shan Guan}
\affiliation{State Key Laboratory of Superlattices and Microstructures, Institute of Semiconductors, Chinese Academy of Sciences, Beijing 100083, China}

\author{Xian-Lei Sheng}
\email{xlsheng@buaa.edu.cn}
\affiliation{Research Laboratory for Quantum Materials, Singapore University of Technology and Design, Singapore 487372, Singapore}
\affiliation{Department of Applied Physics, Key Laboratory of Micro-nano Measurement-Manipulation and Physics (Ministry of Education), Beihang University, Beijing 100191, China}

\author{Shengyuan A. Yang}
\email{shengyuan\_yang@sutd.edu.sg}
\affiliation{Research Laboratory for Quantum Materials, Singapore University of Technology and Design, Singapore 487372, Singapore}

%\date{\today}
\begin{abstract}
    Two-dimensional (2D) materials with nodal-loop band crossing have been attracting great research interest. However, it remains a challenge to find 2D nodal loops that are robust against spin-orbit coupling (SOC) and realized in magnetic states. Here, based on first-principles calculations and theoretical analysis, we predict that monolayer MnN is a 2D nodal-loop half metal with fully spin polarized nodal loops. We show that monolayer MnN has a ferromagnetic ground state with out-of-plane magnetization. Its band structure shows half metallicity with three low-energy bands belonging to the same spin channel. The crossing between these bands forms two concentric nodal loops centered around the $\Gamma$ point near the Fermi level. Remarkably, the nodal loops and their spin polarization are robust under SOC, due to the protection of a mirror symmetry. We construct an effective model to characterize the fully polarized emergent nodal-loop fermions. We also find that a uniaxial strain can induce a loop transformation from a localized single loop circling around $\Gamma$ to a pair of extended loops penetrating the Brillouin zone.
\end{abstract}

%%% pacs
%%% 71.20.-b	Electron density of states and band structure of crystalline solids
%%% 73.20.-r	Electron states at surfaces and interfaces
%%% 31.15.A-	Ab initio calculations

%\pacs{71.20.-b, 73.20.-r, 31.15.A-}
\maketitle

\section{Introduction}

Two-dimensional (2D) materials have been attracting tremendous interest since the discovery of graphene~\cite{novoselov2004,butler2013progress,bhimanapati2015recent,Novoselov2016}. Many remarkable physical properties of graphene are connect to its topological band structure, namely, the conduction and valence bands of graphene linearly cross at two isolated Fermi points in the Brillouin zone (BZ), which makes the low-energy electrons resemble 2D massless Dirac fermions~\cite{neto2009}. With the rapid advance in experimental techniques, many new 2D materials have been fabricated in recent years~\cite{butler2013progress,bhimanapati2015recent,Novoselov2016}, and inspired by graphene, there is great interest to explore 2D materials with nontrivial band topology.

In 2D, topological band crossings can take the form of 0D nodal points and also 1D nodal loops. Several 2D nodal-loop materials have been proposed in theory~\cite{Jin2017,li2018nonsymmorphic,zhou2018coexistence,zhong2019two,Wu2019}, and nodal loops in monolayer Cu$_{2}$Si and monolayer CuSe have been confirmed in recent experiments~\cite{feng2017experimental,gao2018epitaxial}. Compared to nodal points, the condition for stabilizing nodal loops in 2D is typically more stringent.

Two factors related to electron spin play important roles in the stability of nodal loops. The first is the spin-orbit coupling (SOC). Without SOC,  spin is a dummy degree of freedom (for nonmagnetic systems), which only introduces a trivial double degeneracy, and the electrons can be regarded as spinless fermions when analyzing the symmetry/topology protection. However, when SOC is included, spin has to be explicitly considered. Particularly, the symmetry groups for electronic states need to be described by double valued representations rather than single valued ones. Because the number of double valued representations for a group is generally less than their single valued counterparts, a band crossing protected in the absence of SOC is usually destroyed when SOC is turned on. For example, the Dirac points in graphene and most other 2D materials are unstable under SOC~\cite{kane2005quantum}. To have a real Dirac point in 2D, a condition involving certain nonsymmorphic symmetries was proposed~\cite{young2015dirac}, and such spin-orbit Dirac point was recently found in monolayer HfGeTe family materials~\cite{guan2017two}. Similarly, most nodal loops in 2D proposed so far are vulnerable against SOC. For example, the nodal loops in monolayer Cu$_{2}$Si and monolayer CuSe are protected by mirror symmetry in the absence of SOC, but they are removed when SOC is included~\cite{feng2017experimental,gao2018epitaxial}. Hence, to find a 2D material with nodal loop robust against SOC is a much more difficult task.

The second factor is the spin polarization associated with magnetic ordering. 2D magnetic materials with high spin polarization are much desired also from the application perspective, because these materials are useful for compact spintronic devices. Nevertheless, magnetic orders break the time reversal symmetry, which also makes the protection of nodal loops more challenging, especially in the presence of SOC. A few 2D magnetic nodal-loop materials were reported till now, including Na$_{2}$CrBi trilayer~\cite{niu2017topological}, monolayer CrAs$_{2}$~\cite{wang2018type}, and single-layer GdAg$_{2}$~\cite{feng2019discovery}. However, in these examples, both spin-up and spin-down states coexist near the Fermi level, which decreases the net spin polarization for the current carriers. It is thus most desirable to have nodal loops formed in a single spin channel around the Fermi level, i.e., to realize a nodal-loop half metal state. In 3D, such a state was recently predicted for Li$_3$(FeO$_3$)$_2$~\cite{chen2019weyl}, which hosts two fully spin-polarized nodal loops. In 2D, such a nodal-loop half metal has yet to be discovered.

In this work, based on first-principles calculations, we reveal monolayer MnN as the first example of a 2D nodal-loop half metal, with fully spin-polarized nodal loops robust against SOC. A previous work by Xu and Zhu~\cite{xu2018two} has demonstrated that monolayer MnN enjoys good dynamic and thermal stability, and it has a rigid ferromagnetic ordering. Here, we show that the electronic band structure of monolayer MnN features two fully spin-polarized nodal loops near the Fermi level. The two loops are centered around the $\Gamma$ point, protected by a mirror symmetry, and more importantly, robust against SOC. We construct an effective model to describe the low-energy electronic states around the nodal loops. Furthermore, we find that moderate uniaxial strain can transform the outer nodal loop into two nodal loops both traversing the BZ. This corresponds to a transition in the loop winding topology, characterized by a $\mathbb{Z}^2$ index. Our work reveals a promising material platform for exploring the fundamental physics of fully spin-polarized 2D nodal-loop fermions, which also holds great potential for nanoscale spintronics applications.

\section{Computational Method}

Our first-principles calculations were based on the density functional theory (DFT), as implemented in the Vienna \emph{ab-initio} simulation package (VASP)~\cite{kresse1993ab,kresse1996g}. The interaction between electrons and ions was modeled using the projector augmented wave method~\cite{blochl1994pe}. The generalized gradient approximation (GGA) parameterized by Perdew, Burke, and Ernzerhof (PBE) was adopted for the exchange-correlation functional~\cite{perdew1996generalized}. To account for the important correlation effect associated with the Mn-$3d$ orbitals, we included the Hubbard $U$ correction via the PBE$+U$ method~\cite{dudarev1998electron}. Several Hubbard $U$ values were tested, which showed qualitatively similar results. Below, we shall present the results with the $U$ value set as 5 eV. A vacuum space of $20$ {\AA} thickness was used to avoid artificial interactions between periodic images. The energy cutoff was set to be $520$ eV for the plane-wave basis. The Monkhorst-Pack $k$-mesh~\cite{monkhorst1976special} with size $25\times25\times1$ was adopted for the BZ sampling. The lattice parameters and the ionic positions were fully optimized until the residual force on each atom was less than $0.005$ eV/{\AA}. The energy convergence criterion was set to be $10^{-6}$ eV.

\section{Structure and Elastic property}

\begin{figure}[t!]
\includegraphics[width=7.8cm]{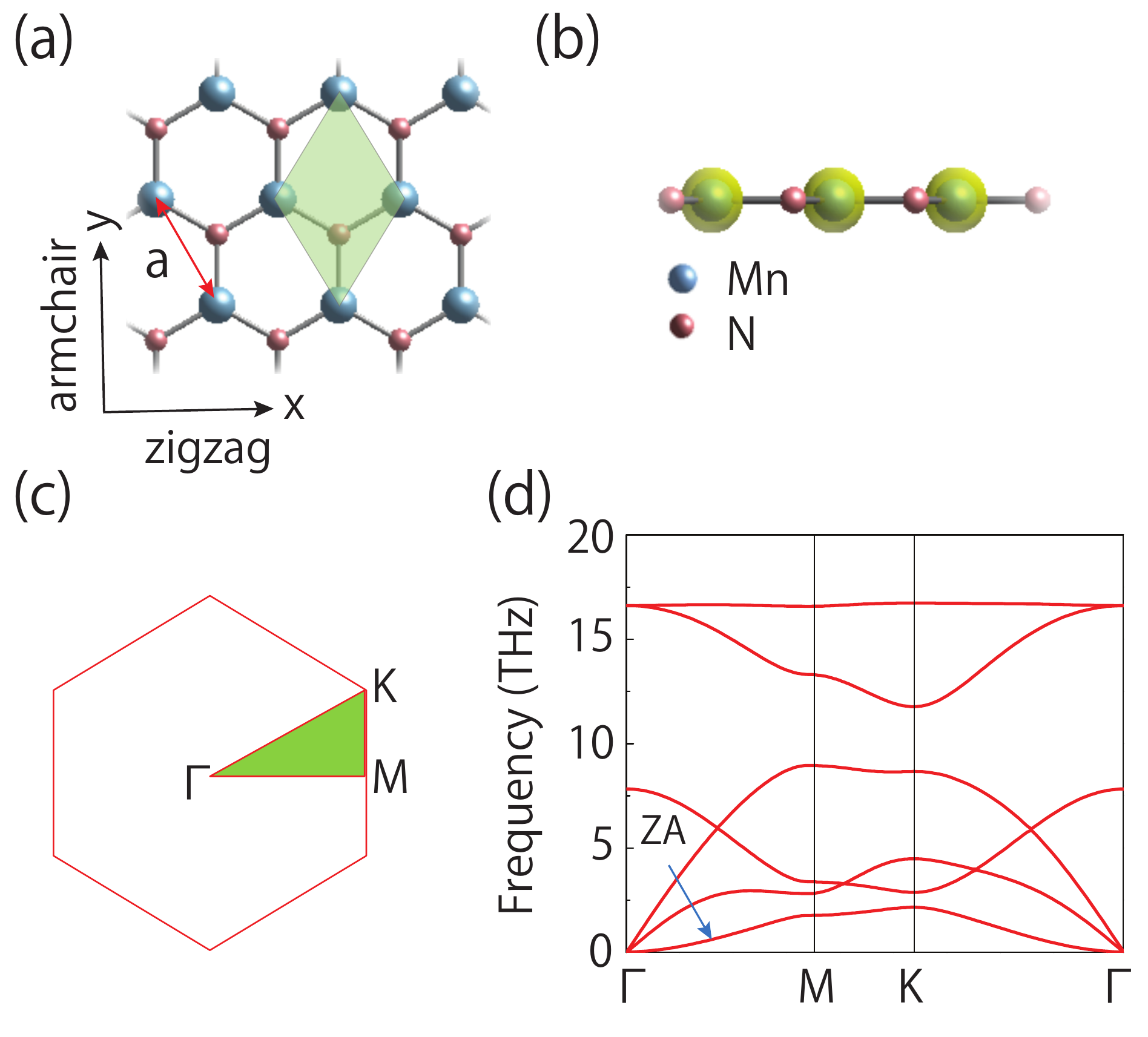}
\caption{(a) Top and (b) side view of the lattice structure for monolayer MnN. The green shaded region in (a) indicates the primitive unit cell. $a$ is the lattice parameter. (b) also shows the spin polarization distribution. The yellow color indicates a net spin-up polarization. (c) Brillouin zone for monolayer MnN. (d) Calculated phonon spectrum. A $6\times6$ supercell is used in the calculation.}
\label{fig1}
\end{figure}

The lattice structure for monolayer MnN is shown in Fig.~\ref{fig1}(a) and \ref{fig1}(b). The material is completely flat with a single atomic layer like graphene. It takes a 2D honeycomb lattice, with Mn and N occupying the A and B sites respectively. The space group is $P\bar{6}m2$, and the point group is $D_{3h}$. The important symmetry here is the horizontal mirror $\mathcal{M}_z$, which will play an important role when discussing the band structure. A primitive unit cell contains one Mn atom and one N atom. From our calculation, the equilibrium lattice constant is $3.388$ \AA\ for the fully relaxed structure, which is in good agreement with the previous work~\cite{xu2018two}. The obtained equilibrium Mn-N bond length is $1.956$ \AA.

The dynamical stability of monolayer MnN can be inferred from its phonon spectrum. As shown in Fig.~\ref{fig1}(d), there is no imaginary frequency phonon mode throughout the BZ, indicating that the material is dynamically stable. Approaching the $\Gamma$ point, there are two linearly dispersing in-plane acoustic branches, and there is also a quadratic out-of-plane acoustic (ZA) branch. The appearance of the quadratic ZA branch is a characteristic feature of 2D materials~\cite{zabel2001phonons,zhu2014coexistence,carrete2016physically}. The sound speed for the longitudinal acoustic phonons in monolayer MnN ($\sim 7.1$ km/s) is smaller than that of graphene ($\sim 21.2$ km/s)~\cite{kaasbjerg2012unraveling} as well as blue phosphorene ($\sim 8.1$ km/s)~\cite{zhu2014semiconducting}, indicating that monolayer MnN should have relatively small in-plane stiffness, as we show below.

\begin{figure}[t!]
\includegraphics[width=8.4cm]{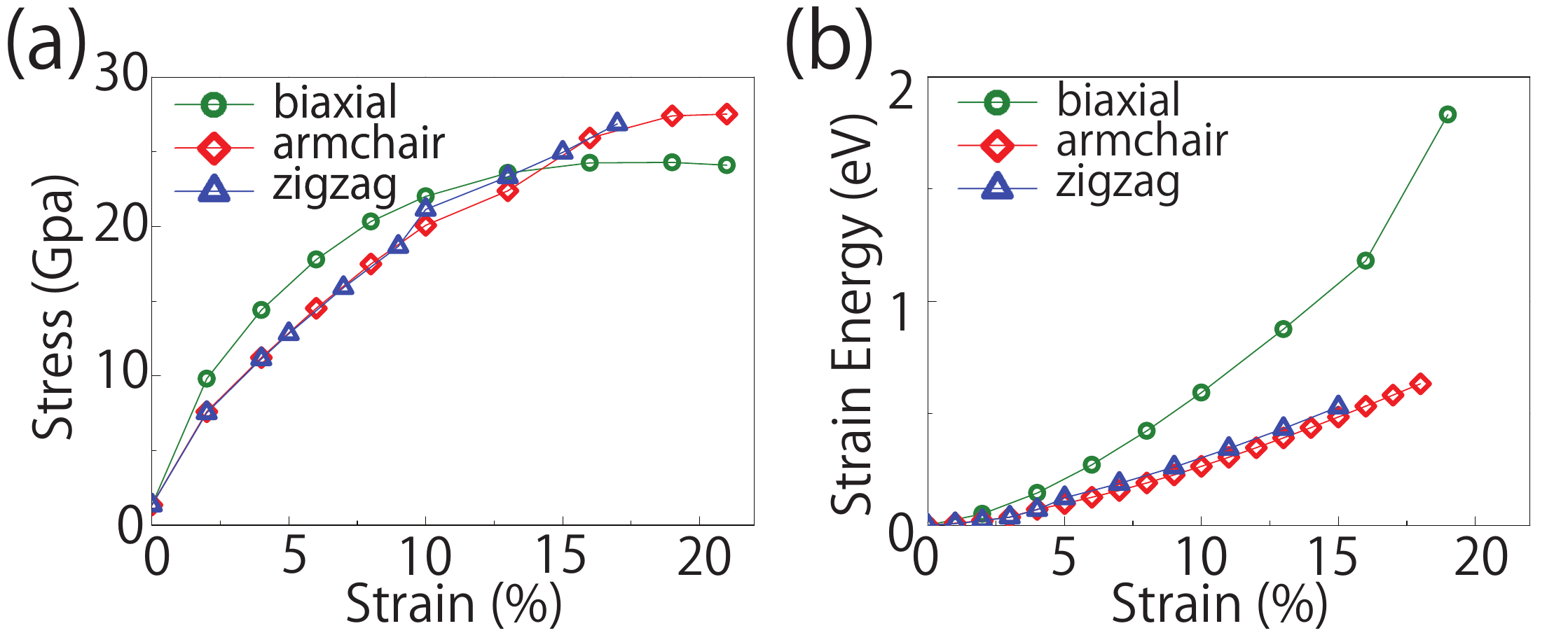}
\caption{(a) Strain-stress relation for monolayer MnN with different types of strain. (b) Strain energy ($E_{s}$) as a function of applied strain. }
\label{figss}
\end{figure}

Next we examine the elastic properties. The calculated strain-stress and strain-energy curves are plotted in Fig.~\ref{figss}. One can observe that
monolayer MnN can sustain a large biaxial strain up to $20\%$, while the critical uniaxial strain can be $\sim 18\%$ for the armchair direction and $\sim 20\%$ for the zigzag direction. These values are comparable to other 2D materials such as graphene~\cite{lee2008measurement}, MoS$_2$~\cite{castellanos2012mechanical,bertolazzi2011stretching}, phosphorene~\cite{peng2014strain}, and C$_2$N~\cite{Guan2015}.
Monolayer MnN remains within the linear elastic regime until about $5\%$ biaxial strain and $8\%$ uniaxial strain. For small deformations, the elastic property of $2$D materials is usually characterized by the in-plane stiffness constant, defined as
\begin{equation}
C=\frac{1}{S_{0}}\frac{\partial^{2}E_{s}}{\partial\varepsilon^{2}},
\end{equation}
where $S_{0}$ is the equilibrium area of the unit cell, $E_{s}$ is the strain energy (i.e., the energy difference between the strained and unstrained systems), and $\varepsilon$ is the in-plane uniaxial strain. From the stain-energy curve in Fig.~\ref{figss}(b), the typical quadratic dependence can be observed at small deformations. The obtained stiffness constants are $65$ N/m and $62$ N/m for strains along the armchair and the zigzag directions, respectively. These values are smaller than other typical 2D materials such as graphene ($\sim 340\pm 40$ N/m)~\cite{lee2008measurement}, MoS$_{2}$ ($\sim 120$ N/m)~\cite{peng2013outstanding}, and BN ($\sim 267$ N/m)~\cite{topsakal2010response}, and are similar to C$_2$N ($\sim 71$ N/m)~\cite{Guan2015} and Mg$_2$C ($\sim 60$ N/m)~\cite{wang2018monolayer}.
This indicates that monolayer MnN is quite flexible, which will facilitate the strain engineering of its physical properties.

\section{Magnetic property}

The $3d$ transition metal elements like Mn often carry magnetic moments due to their partially filled $3d$ shell. The materials containing Mn hence often exhibit magnetic ordering, which is the case for monolayer MnN. In our calculation, we
compare the energies of different magnetic configurations: nonmagnetic (NM), ferromagnetic (FM), and antiferromagnetic (AFM), to determine the ground state of monolayer MnN.
It is found that the FM state is energetically favored, while the NM sate and the AFM state are respectively $4.68$ and $0.39$ eV higher than the FM state per unit cell. In the FM state, the magnetic moment is mainly distributed on the Mn sites [see Fig.~\ref{fig1}(b)], with $\sim 3.99\mu_B$ per Mn atom.

To pin down the easy axis for the FM state, the magnetic anisotropy energy calculation was performed by scanning different orientations for the magnetic moment with SOC included. The result shows that the out-of-plane (along $z$) configuration is most energetically favorable, which is about 0.4 meV lower in energy than the in-plane configuration. Thus, the ground state magnetic configuration for monolayer MnN is an FM state with magnetization along the out-of-plane $z$ direction.

We have also estimated the Curie temperature $T_C$ for the FM state, by using the Monte Carlo simulation based on a classical Heisenberg-like spin model~\cite{evans2014atomistic}:
\begin{equation}
  H=-J\sum_{\langle i,j\rangle}{\bm S}^{i}\cdot {\bm S}^{j}-K\sum_i (S_z^i)^{2},
\end{equation}
where superscripts $i$ and $j$ label the Mn sites, $\langle i,j\rangle$ indicates nearest neighboring sites, $J$ is the FM exchange coupling strength, and $K$ represents the magnetic anisotropy strength. The calculated Curie temperature for monolayer MnN is around 200 K, which is comparable with other 2D ferromagnetic materials discovered to date, such as  Cr$_2$Ge$_2$Te$_6$ atomic layers ($\sim$66 K)~\cite{Gong2017}, transition metal halides (23-128 K)~\cite{Huang2017,kulish2017single}, and Fe$_{3}$GeTe$_{2}$ ($\sim$130 K)~\cite{fei2018two}. %{\color{red} there is another experimental one.}

\section{2D Nodal-loop half metal}

\begin{figure}[t!]
\includegraphics[width=8.2cm]{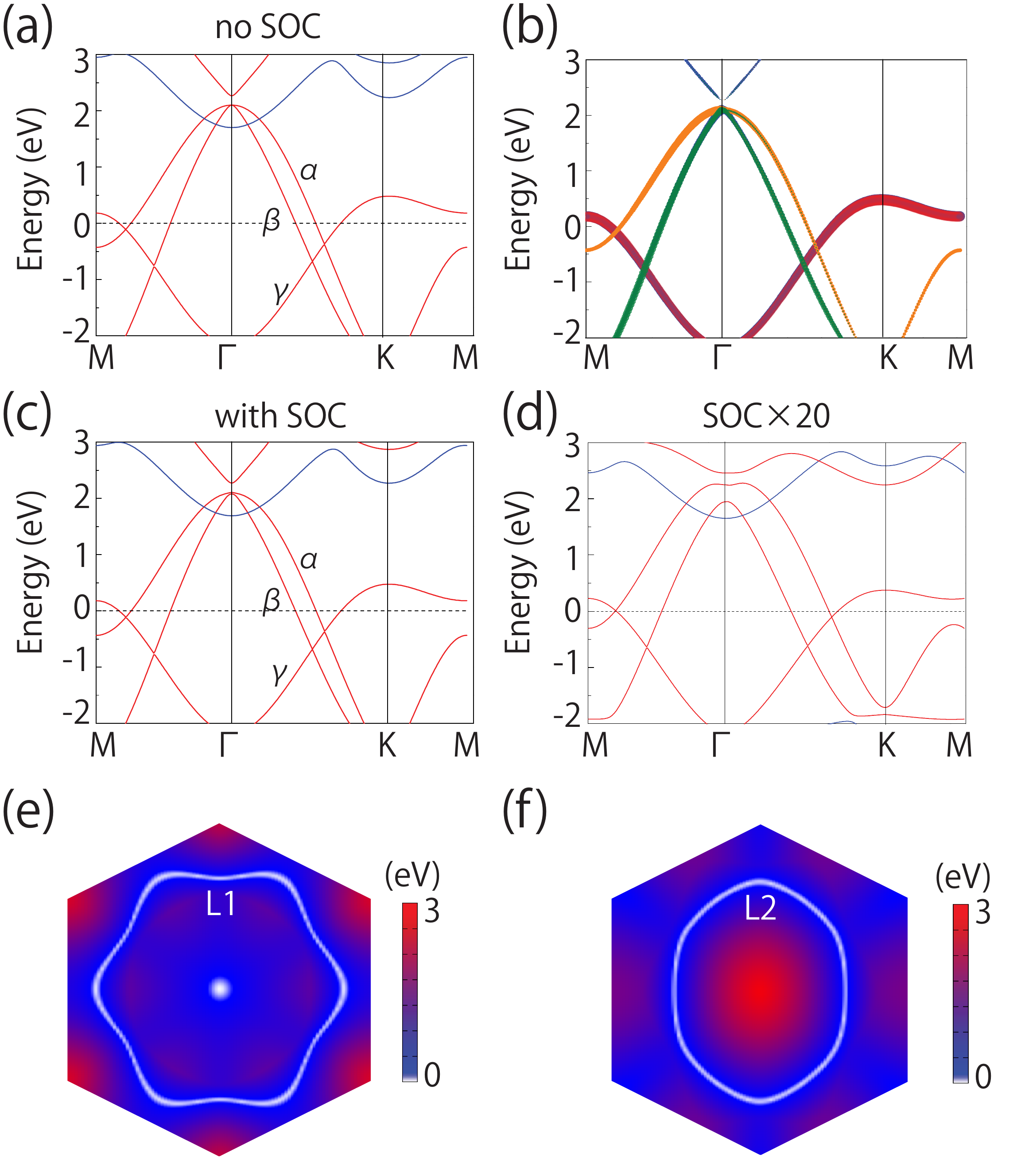}
\caption{(a) Calculated electronic band structure for monolayer MnN in the absence of SOC. (b) shows the projection of the bands in (a) onto atomic orbitals, including N $p_{x}/p_{y}$ orbital (blue), N $p_{z}$ orbital (red), Mn $d_{xy}$ orbital (yellow), and Mn $d_{x^{2}-y^{2}}$ orbital (green). (c) Calculated band structure with SOC. (d) Band structure with SOC artificially enhanced by 20 times in the calculation. (e) and (f) show the shapes of the two nodal loops obtained by DFT calculation. The color map indicates the local gap between two crossing bands. In figures (a), (c), and (d), the red and blue colors indicate bands with fully polarized spin-up and spin-down states, respectively, showing that the two loops are fully spin polarized. }
\label{Band}
\end{figure}

The most interesting property of monolayer MnN lies in its electronic band structure. Let's first examine the band structure in the absence of SOC, as shown in Fig.~\ref{Band}(a). Two salient features can be observed. First, the ground state of monolayer MnN is a half metal. The bands for majority spin (denoted as spin up) display a metallic behavior, whereas the bands for minority spin (denoted as spin down) have a large energy gap about $4.5$ eV. The low-energy electrons near the Fermi level are therefore fully spin polarized.  Second, several linear band crossings are observed for the majority spin bands close to the Fermi level. Among these low-energy bands, one is electron-like (labeled as $\gamma$ band) and the other two are hole-like (labeled as $\alpha$ and $\beta$ bands). From orbital projections in Fig.~\ref{Band}(b), we find that the $\alpha$ and $\beta$ bands are mainly contributed by the Mn ($d_{xy}$, $d_{x^{2}-y^{2}}$) orbitals, while the $\gamma$ band is mainly from the N $p_{z}$ orbital. The band crossings occur between the $\gamma$ band and the other two bands. A careful scan of the band structure shows that these crossing points are not isolated, instead, they actually belong to two concentric nodal loops centered around the $\Gamma$ point, as shown in Fig.~\ref{Band}(e) and \ref{Band}(f).
We label the outer loop and the inner loop as $L_1$ and $L_2$, respectively.

Remarkably, the two nodal loops persist when SOC is turned on. Figure~\ref{Band}(c) shows the band structure with SOC included. The result is similar to that in Fig.~\ref{Band}(a). To show that the nodal loops are indeed preserved under SOC, we artificially increase the SOC strength by 20 times in the DFT calculation, and plot the resulting band structure in Fig.~\ref{Band}(d). It clearly shows that the crossings are robust against SOC. What protects the nodal loops? By analyzing the symmetry properties of the three bands, we find that the two loops are protected by the mirror symmetry $\mathcal{M}_z$. Because the ground state magnetization is along the out-of-plane $z$ direction, $\mathcal{M}_z$ is preserved in the FM state. The two hole-like bands $\alpha$ and $\beta$ have the $\mathcal{M}_z$ eigenvalue $+i$, whereas the electron-like band $\gamma$ has the opposite $\mathcal{M}_z$ eigenvalue $-i$. Therefore, $\gamma$ band must cross $\alpha$ and $\beta$ bands without hybridization, and the resulting two loops are protected as long as the mirror symmetry is maintained. Furthermore, because of the $\mathcal{M}_z$ symmetry, the eigenstates must be spin polarized along $z$, i.e., the spin-up and spin-down bands do not hybridize and each band remains fully spin polarized. Thus, monolayer MnN indeed provides a 2D nodal-loop half metal with fully spin-polarized nodal loops robust under SOC.

A nodal loop can be classified as type-I, type-II~\cite{li2017type}, or hybrid type~\cite{li2017type,zhang2018hybrid}, based on the type
of dispersion around the loop. A type-I (type-II) nodal loop consists solely of type-I (type-II) nodal points, while a hybrid loop consists of both types~\cite{li2017type,zhang2018hybrid}. The two nodal loops in monolayer MnN belong to type-I, because they each is formed by the crossing between an electron-like band and a hole-like band.

To characterize the emergent nodal-loop fermions, we construct an effective $k\cdot p$ model for the low-energy band structure.
We choose the three states at the $\Gamma$ point as the basis. We first consider the case without SOC. The state of the $\alpha$ band corresponds to the $A''_{2}$ irreducible representation for the $D_{3h}$ point group, while the two degenerate states of $\beta$ and $\gamma$ bands correspond to the $E'$ representation. The model should respect the following symmetries: the threefold rotation $C_{3z}$, the twofold rotation $C_{2x}$, and $\mathcal{M}_{z}$. Expanding up to the $k$ quadratic order, we find that the effective Hamiltonian takes the form of
\begin{equation}\label{H0}
\begin{split}
    &\mathcal{H}_{0}(\bm k)= \\
    &\left[\begin{array}{ccc}
    M_0+A k_{x}^{2} +B k_{y}^{2}& (B-A)k_{x}k_{y} & 0 \\
    (B-A)k_{x}k_{y}&  M_0+B k_{x}^{2} +A k_{y}^{2}& 0 \\
    0 & 0 & M_{1}+C k^{2}
    \end{array}\right],
\end{split}
\end{equation}
where $M_i$, $A$, $B$, and $C$ are real valued parameters which can be obtained from fitting the DFT band structure. Note that the matrix element between the $\alpha$ ($\beta$) and $\gamma$ bands vanishes identically, because the two bands have opposite $\mathcal{M}_{z}$ eigenvalues.
Including SOC gives additional contribution, which can be treated as a perturbation due the relatively weak SOC strength. We find that the SOC term expanded to leading order takes the following form
\begin{equation}\label{SOC}
\begin{split}
    \mathcal{H}_\text{SOC}(\bm k)=
    \left[\begin{array}{ccc}
    0 & iD & 0 \\
    -iD &  0 & 0 \\
    0 & 0 & 0
    \end{array}\right],
\end{split}
\end{equation}
with some real valued parameter $D$. Obviously, the SOC term does not affect the presence of the two nodal loops. Its main effect is that the degeneracy for the $E'$ states is lifted. This effective model fits the DFT band structure very well (see the Supplemental Material~\cite{supp}).

\begin{figure}[t!]
\includegraphics[width=8.2cm]{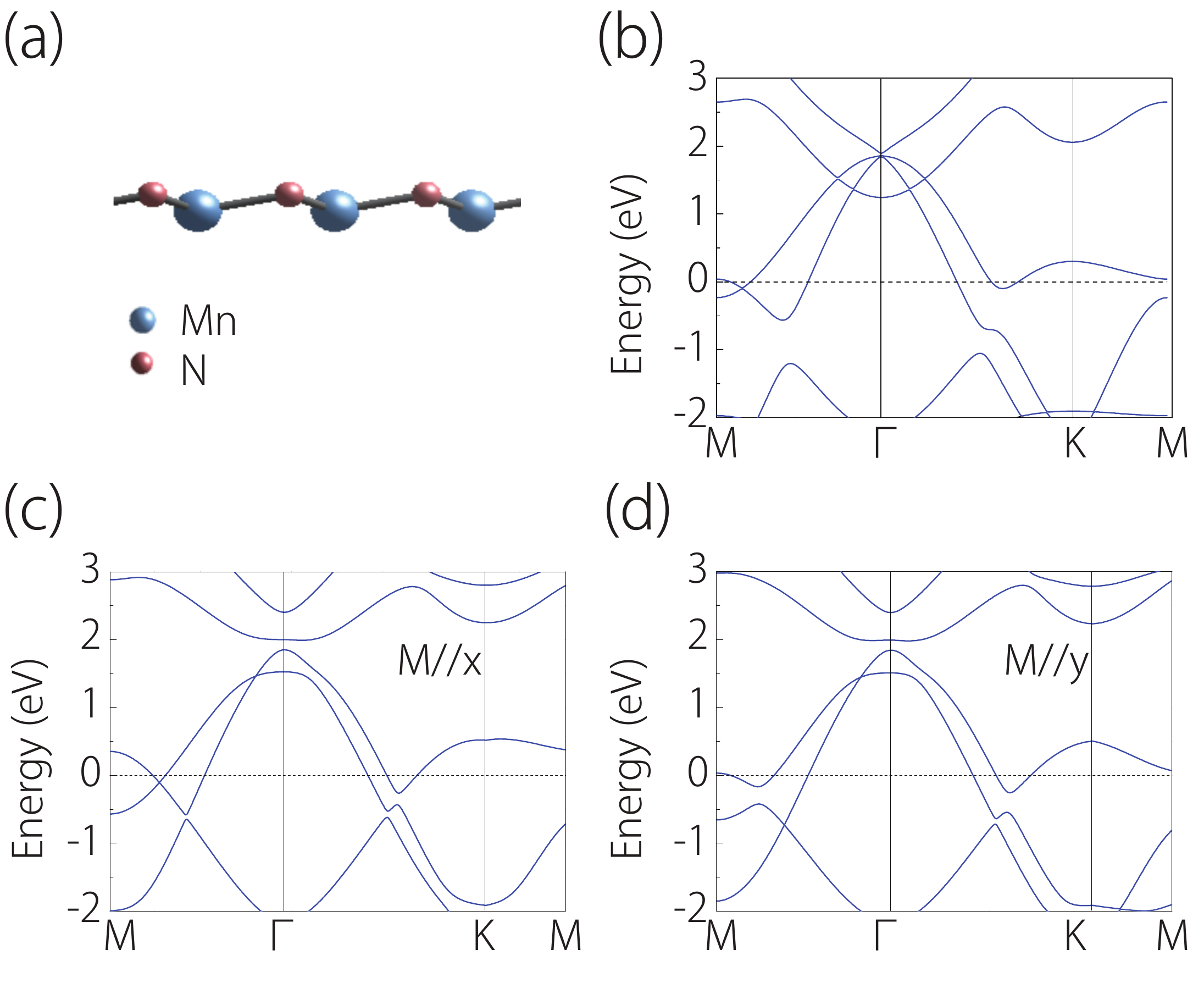}
\caption{(a) shows a lattice distortion that breaks the horizontal mirror reflection symmetry. Here, the monolayer shows a buckling with the Mn atoms shifted up relative to the N atoms. (b) Calculated band structure (including SOC) for the distorted structure in (a) when the shift is 0.2 \AA. (c) and (d) present the band structures (including SOC) for \emph{undistorted} lattice but with magnetic moments oriented in-plane along $x$ and $y$ directions, respectively.}
\label{figMB}
\end{figure}

Breaking the mirror $\mathcal{M}_{z}$ would generally destroy the nodal loops. This can be done, e.g., by artificially shifting the Mn atoms along $z$ relative to the N atoms, making a buckled honeycomb lattice [see Fig.~\ref{figMB}(a)], or by orienting the magnetization direction in-plane. The corresponding band structure results in Fig.~\ref{figMB}(b-d) show that the two loops are indeed gapped out when $\mathcal{M}_{z}$ is broken.

\section{Strain-induced nodal loop transformation}

\begin{figure}[t!]
\includegraphics[width=8.2cm]{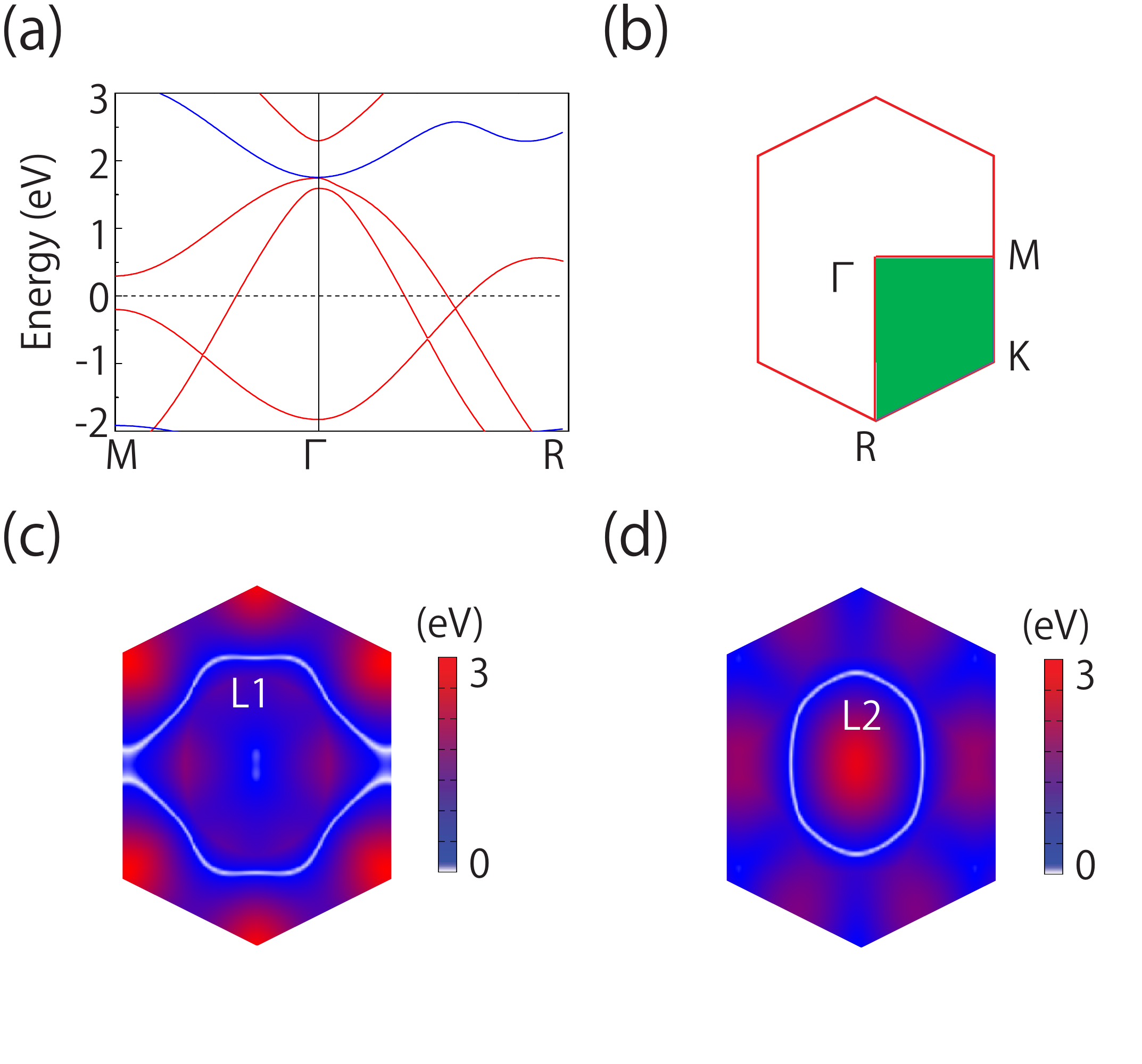}
\caption{(a) Calculated band structure for monolayer MnN with 7\% uniaxial strain along the armchair direction. The red and blue colors represent spin-up and spin-down bands, respectively. (b) shows the corresponding Brillouin zone. (c) and (d) show the shapes of the corresponding nodal loops obtained from DFT calculation. $L_1$ is transformed into two extended loops traversing the Brillouin zone. The color map indicates the local gap between two crossing bands.}
\label{figBand}
\end{figure}

As discussed before, monolayer MnN enjoys good flexibility, hence its physical property should be readily tunable by strain.
Here, we focus on the strain effect on the nodal loops. Since the two loops are protected by the horizontal mirror, they must be preserved under a variety of in-plane strains, such as the biaxial, uniaxial, and shear strains. Nevertheless, the shape of the loop may be changed by strain.
For example, we consider the uniaxial strain along the armchair direction. Our result in Fig.~\ref{figss}(a) has shown that the critical strain is above $18\%$. In
Fig.~\ref{figBand}(a), we show the band structure result for the 7\% strain. One observes that while the loop $L_2$ remains a closed circle around the $\Gamma$ point, the crossing point on $\Gamma$-$M$ for loop $L_1$ disappears, because the ordering between $\alpha$ and $\gamma$ bands at $M$ is switched.
Accordingly, the $L_1$ loop is transformed from a localized loop circling around $\Gamma$ into two extended loops traversing the BZ, as shown in Fig.~\ref{figBand}(c).

This transformation corresponds to a change in the loop winding topology defined on the BZ. In Ref.~\cite{li2017type}, it has been shown that loops can be characterized by its winding pattern in the BZ, captured by the fundamental homotopy group for the BZ. For 2D, the BZ is a two-torus. Since $\pi_{1}(\mathbb{T}^{2})=\mathbb{Z\times\mathbb{Z}}$, the topology is characterized by two integers, each standing for the number of times the loop winds around the BZ in a particular direction. For the current case, the original $L_1$ without strain is characterized by an index $(0,0)$, whereas the two loops after transformation [in Fig.~\ref{figBand}(c)] are characterized by $(0, \pm 1)$. Their topological distinction can be understood in the following way. The loop with index $(0,0)$ can be continuously deformed into a point and annihilate, whereas a single loop with index $(0,1)$ cannot. Similar transformation was recently reported in the 2D honeycomb borophene oxide~\cite{zhong2019two}. The monolayer MnN offers another platform for studying such interesting nodal loop transformation.

\section{Discussion and Conclusion}

We have a few remarks before closing. First,
the finding reported here is significant, because it reveals the first example of a 2D nodal-loop half metal and the loops are robust against SOC. Almost all the nodal loops in 2D reported so far are not robust under SOC. 2D magnetic nodal-loop materials are also very rare. It is very fortunate that the nodal loop, its robustness against SOC, and half metallicity can occur in the same material. Besides, the material band structure has the following two additional advantages: (i) The low-energy bands for monolayer MnN are fairly simple and clean, without other extraneous bands; and (ii) the linear dispersion range is also quite large ($>1$ eV).

Second, as the nodal loops here are protected by symmetry, their existence is not sensitive to the first-principles calculation methods. For example, we have verified that qualitative features of the band structure remain the same when using the hybrid functional (HSE06) approach. The result can be found in the Supplemental Material~\cite{supp}.

Finally, we remark on several experimental aspects. It has been suggested that monolayer MnN can be synthesized by a spontaneous graphitic conversion from the ultrathin (111)-oriented cubic MnN in experiment~\cite{sorokin2014spontaneous}. The bottom-up approaches such as the molecular beam epitaxy (MBE) may also offer an alternative. It has been demonstrated that 2D monolayer CuSe with similar structure can be synthesized via the MBE approach with methane on the Cu(111) substrate~\cite{lin2017intrinsically}. The method is shown to be very versatile, and can be used as a general strategy for fabricating other 2D materials.
The nodal loops can be directly imaged by using the angle-resolved photoemission spectroscopy (ARPES), as recently demonstrated for monolayer Cu$_2$Si and CuSe~\cite{feng2017experimental,gao2018epitaxial}. The nodal loops may also produce features in the quasiparticle interference pattern which can be probed by the scanning tunneling microscopy experiment~\cite{zheng2016atomic,zhu2018quasiparticle}. Techniques for applying strain on 2D materials have been well developed in recent years. For example, strain can be applied by using a beam bending apparatus~\cite{conley2013bandgap} or by using an atomic force microscope tip~\cite{lee2008measurement}. By using a stretchable substrate, controllable strains $>10\%$ have been demonstrated for graphene~\cite{kim2009large}.

In conclusion, we have revealed monolayer MnN as a 2D nodal-loop half metal. We show that the material has good stability, can sustain large strains, and has excellent flexibility. The material has a FM ground state with an estimated Curie temperature about 200 K.
Most interestingly, its band structure shows a half metal and features two concentric nodal loops near the Fermi level in a single spin channel. These nodal loops are protected by the mirror symmetry and are robust against SOC.
We construct an effective model to characterize the emergent nodal-loop fermions. In addition, we find that a moderate uniaxial strain can generate an interesting loop topology transformation, from a single loop circling around $\Gamma$ to two loops traversing the BZ.
Our work thus provides a promising platform to investigate the intriguing properties of nodal-loop fermions with time-reversal breaking. We hope that our theoretical work will facilitate the experimental studies on this new 2D material towards both fundamental discoveries and potential spintronics applications.

\begin{acknowledgements}
The authors thank Hongming Weng, Ran Shen, and D. L. Deng for valuable discussions. This work is supported by the Singapore Ministry of Education Academic Research Fund Tier 2 (MOE2015-T2-2-144 and MOE2017-T2-2-108). We acknowledge computational support from the Texas Advanced Computing Center and National Supercomputing Centre Singapore.
\end{acknowledgements}

%\begin{references}
%\end{references}

\bibliography{MnN_ref}

%\newpage

\end{document}